\documentstyle[12pt,aasms4]{article}

\received{June 29 2001}
\accepted{December 00 2001}

\begin{document}

\title{Reexamination of the Galaxy Formation-Regulated Gas Evolution
Model in Groups and Clusters}

\author{Xiang-Ping Wu and Yan-Jie Xue}

\affil{National Astronomical Observatories, Chinese Academy
                 of Sciences, Beijing 100012; China}

\begin{abstract}
As an alternative explanation of the entropy excess and the steepening
of the X-ray luminosity-temperature relation in groups and clusters,
the galaxy formation-regulated gas evolution (GG) model proposed 
recently by Bryan makes an attempt to incorporate the formation of 
galaxies into the evolution of gas without additional heating
by nongravitational processes. 
This seems to provide a unified scheme for our understanding of
the structures and evolution of both galaxies and gas 
in groups and clusters.
In this paper, we present an extensive comparison of the X-ray  
properties of groups and clusters predicted by the GG model and 
those revealed by current X-ray observations, 
using various large data sources in the literature
and also taking the observational selection effects into account.
These include an independent check of the fundamental working hypothesis of 
the GG model, i.e., galaxy formation was less
efficient in rich clusters than in groups,
a new test of the radial gas distributions revealed 
by both the gas mass fraction and the X-ray surface brightness profiles,
and an reexamination of the X-ray luminosity-temperature and 
entropy-temperature relations. 
In particular, it shows that the overall X-ray surface brightness profiles
predicted by the GG model are very similar in shape, insensitive to
the X-ray temperature, and the shallower X-ray surface brightness profiles
seen at low-temperature systems may arise from the current observational  
selection effect. This can be used as the simplest approach to distinguishing 
between the GG model and the preheating scenario. The latter yields
an intrinsically shallower gas distribution in groups than in rich clusters.
\end{abstract}


\section{Introduction}

It has been realized for a decade that the ratio of gas mass to stellar
mass in groups and clusters of galaxies is positively correlated
with the X-ray gas temperature. This suggests a simple scenario
in which galaxy formation was more efficient in groups than 
in clusters  (David et al. 1990). As a consequence, 
the stellar (gas) mass fraction should decrease  (increase) from groups 
to clusters, resulting in the increasing mass-to-light ratio and 
X-ray luminosity with X-ray temperature. It seems that 
these predictions have been well justified by the optical measurements of 
the mass-to-light ratios (e.g. Girardi et al. 2000)
and the X-ray determinations of the gas fractions in clusters 
(e.g. Mohr, Mathiesen \& Evrard 1999; MME), and in particular by 
the discovery of the steepening of the X-ray luminosity 
($L_{\rm x}$)-temperature ($T$) relation 
for groups and clusters (Henry \& Arnaud 1991; Edge \& Stewart 1991; 
David et al. 1993; Wu, Xue \& Fang 1999; Xue \& Wu 2000 and references
therein).  Nevertheless, despite these apparent successes,  such a simple
scenario has received less attention over the past years than the prevailing 
`preheating' hypothesis. The latter assumes that the gas was preheated by 
nongravitational processes such as supernovae and AGNs before collapsing
into groups and clusters, which raises the gas entropy and makes the 
warm/hot gas harder to compress (Kaiser 1991). 
Indeed, the nongravitational heating mechanism can significantly 
reduce the gas content bound within the virial radii of  
groups and poor clusters, providing a 
satisfactory explanation of the observational facts listed above
(e.g. Cavaliere, Menci \& Tozzi 1997; 1999; Bower et al. 2001). 
In particular, the subsequent detection of the entropy excess 
in the centers of groups and clusters by 
Ponman, Cannon \& Navarro (1999) seems to give 
a convincing  support to the preheating hypothesis. 
Yet, the main difficulty with the preheating model is the unreasonably 
high efficiency of energy injection into the intragroup/intracluster gas 
from supernovae (Wu, Fabian \& Nulsen 2000; Bower et al. 2001), although
additional heating from AGNs might help to alleviate the discrepancy.
Furthermore, a uniform preheating of the cosmic baryons to 
a temperature of $\sim10^6$ at high redshifts may make the Ly$\alpha$ 
forest ($T\sim10^4$ K) to disappear.

By inheriting the observational evidence that the efficiency of galaxy 
formation was higher in groups than in clusters, Bryan (2000) proposed
that the lowest entropy gas in groups and clusters should be responsible 
for the formation of stars. Namely, the gas in the central regions
of groups and clusters was converted into stars in the early phase of
structure formation, and the central gas cavity was refilled with the 
remaining gas distributed originally at large radii. 
This latter process brought the higher entropy gas inward and thus
raised the central entropy of the gas, providing an alternative 
explanation of the existence of the entropy floor in the centers of 
groups and clusters reported by Ponman et al. (1999).  
The most important issue behind this simple scenario, we believe, is 
that the distribution and evolution of the gas  
are incorporated with the formation of the galaxies so that 
the optical/X-ray properties of the gas and galaxies in groups
and clusters can be naturally explained within a unified scheme
and without invoking additional heating processes. 
This galaxy formation-regulated gas evolution  (GG) model 
in groups and clusters has several unambiguous predictions that can  
be easily tested with current available data or future
observations. For example, (1)while the stellar and gas mass fractions 
vary from groups to clusters, the total baryon (stellar mass + gas mass) 
fraction in any group or cluster remains a universal value; 
(2)The mass-to-light ratio is an increasing function of X-ray temperature; 
(3)Less massive
systems (e.g. groups) contain less gas and hence have lower X-ray emission,
which is the key issue for the reproductions of the observed 
$L_{\rm x}$-$T$ relation and entropy distribution;
(4)Radial profile of the stellar mass fraction should demonstrate a 
decreasing trend towards large radii, and an opposite situation  
occurs for that of the gas mass fraction.  Of course, whether or 
not these properties are in quantitative agreement with observations 
deserves  further investigation.

In the pioneering work of Bryan (2000), an overall agreement 
between the predicted and observed $L_{\rm x}$-$T$ relations and 
entropy distributions has been essentially found, although at 
low temperature $T<1$ keV, the observed $L_{\rm x}$-$T$ relation 
falls below the theoretical prediction.  Motivated by these successes and
their profound implications for our understanding of the formation
and evolution of the galaxies and gas in groups and clusters, 
we wish to conduct an extensive comparison of the theoretically 
predicted  properties of the intragroup/intracluster 
gas by the GG model with those revealed by current X-ray observations. 
These include not only the global dependence of
the X-ray luminosity, entropy and gas fraction upon the X-ray temperature $T$, 
but also the internal structures of the intragroup/intracluster gas such as  
the radial distribution of gas fraction, the X-ray surface 
brightness profile (slope and core radius parameters), etc. 
Moreover, some selection effects of current X-ray observations 
(e.g. the finite extensions of X-ray surface brightness profiles 
due to background noise,  the spectral fitting temperature, etc.) 
will be taken into account. It is hoped that our comparison
would be useful for an eventually conclusive test of the GG model.
Throughout this paper we assume a flat cosmological  
model ($\Lambda$CDM)
of $\Omega_{\rm M}=0.3$, $\Omega_{\Lambda}=0.7$ and $h=0.65$.

\section{The Model}

\subsection{Reexamination of the working hypothesis}

We begin with a reexamination of the fundamental relationship between 
the stellar mass fraction $f_{\rm star}$ and X-ray temperature $T$ 
of groups and clusters. Based on three independent studies of 33 groups
and clusters in the literature, Bryan (2000) derived the following 
empirical formula, regardless of the large scatter of the data points
especially for low-temperature groups,
\begin{equation}
f_{\rm star}=(0.042\pm0.003)
             \left(\frac{kT}{10\;{\rm keV}}\right)^{-0.35\pm0.06},
\end{equation}
where the quoted errors are $68\%$ confidence limits.
The lack of a large sample of optical/X-ray groups and clusters
in which both stellar mass and gas mass are reliably determined
out to virial radii $r_{\rm vir}$ hinders us from directly reexamining the 
validity of this relation. Nevertheless, there exist several indirect 
ways that may allow us to test the consistency between equation (1)
and other independent measurements.

The GG model 
admits the universality of the baryon fraction $f_b$ in different 
groups and clusters: $f_b=f_{\rm star}+f_{\rm gas}$. For the $\Lambda$CDM model
and $\Omega_b=0.020h^{-2}$, we have $f_b=0.16$. Therefore, the measurement
of the gas fraction $f_{\rm gas}$ in different groups and clusters permits 
an indirect test of the dependence of the stellar mass fraction upon 
the X-ray temperature. The major difficulty of this exercise is the 
determination of total group/cluster mass $M_{\rm vir}$,  in which   
one often needs to extrapolate the available X-ray data
within finite detection aperture to virial radii $r_{\rm vir}$, 
along with the assumption of hydrostatic equilibrium. Such an extrapolation
may be safely applicable only to X-ray luminous and rich clusters.   
For this reason, 
we extract a sample of 45 nearby ($z<0.1$), X-ray luminous clusters
from MME and Ettori \& Fabian (1999; EF).
The restriction on cluster redshifts is consistent with the three
samples used by Bryan (2000), and also gets rid of the influence of 
cosmological parameters other than the Hubble constant.
These authors have explicitly derived the gas fraction within $r_{500}$ 
where the overdensity $\Delta$ of the cluster dark matter with respect to 
the critical density of the universe is $\Delta=500$.  $r_{500}$ can be 
approximately regarded as the virial radius if we notice that the asymptotic
gas fractions of rich clusters at large radii remain roughly constant
for the conventional isothermal $\beta$ model with $\beta=2/3$,
although the virial radius of a dynamically-relaxed system in the $\Lambda$CDM 
model corresponds to $\Delta\approx100$. We display in Figure 1
the observationally determined gas fraction and the expectation 
of $f_{\rm gas}=0.16-0.042(kT/10\;{\rm keV})^{-0.35}$ from
the empirical formula equation (1) by Bryan (2000). It is apparent
that the two results are essentially consistent with each other.

\placefigure{fig1}

Another independent check of equation (1) is to examine the 
dependence of the mass-to-light ratio $M/L$ upon the X-ray temperature. 
It follows that the  mass-to-light ratio $M/L$ can be rewritten as 
$M/L=(M/M_{\rm star})(M_{\rm star}/L)=\Upsilon f_{\rm star}^{-1}$,
where $M_{\rm star}$ is the total stellar mass of groups/clusters
and $\Upsilon\equiv M_{\rm star}/L$,  while
observations suggest that $\Upsilon\approx6.5\Upsilon_{\odot}$ for the 
dominant population of galaxies in clusters, ellipticals 
(see Fukugita, Hogan \& Peebles 1998). As a result, the mass-to-light
ratio should increase with the X-ray temperature according to equation (1).
Now, instead of examining the dependence of  $M/L$ upon $T$, 
we would like to derive the $f_{\rm star}$-$T$ relation using
the best-fit relation between cluster virial mass and optical luminosity 
found by Girardi et al. (2000):
\begin{equation}
\left(\frac{M_{\rm vir}}{M_{\odot}h^{-1}}\right)=
       10^c\left(\frac{L}{L_{\odot}h^{-2}}\right)^d,
\end{equation}
where $c=-1.41$ and $d=1.32$ for the fore/background 
correction based on mean galaxy counts. This yields
\begin{equation}
f_{\rm star}=10^{-c/d} 
		 \left(\frac{\Upsilon}{h\Upsilon_{\odot}}\right)
             \left(\frac{M_{\rm vir}}{h^{-1}M_{\odot}}\right)^{-1+1/d}.
\end{equation}
Employing the cosmic virial theorem at $z=0$ (e.g. Bryan \& Norman 1998)
\begin{equation}
\left(\frac{M_{\rm vir}}{10^{15}M_{\odot}}\right)=
              \frac{1}{h\sqrt{\Delta}}
              \left(\frac{kT}{1.39\;{\rm keV}}\right)^{3/2},
\end{equation}
we have 
\begin{equation}
f_{\rm star}=  \left(1.93\times10^{16} \right)^{-1+1/d}
   10^{-c/d} \Delta^{(1-1/d)/2}
   \left(\frac{\Upsilon}{h\;\Upsilon_{\odot}}\right)
   \left(\frac{kT}{10\;{\rm keV}}\right)^{3(-1+1/d)/2}.
\end{equation}
Finally, 
\begin{equation}
f_{\rm star}=0.025\left(\frac{kT}{10\;{\rm keV}}\right)^{-0.36},
\end{equation}
in which we have adopted $\Upsilon=6.5\Upsilon_{\odot}$ and
$\Delta=200$.
Actually, one can arrive at the essentially same conclusion using 
the empirical $M_{\rm vir}$-$T$ relationship for clusters instead of
the cosmic virial theorem equation (4). For example, the best-fit
$M_{\rm 200}$-$T$ relation based on the spatially resolved temperature
profiles in the estimates of X-ray cluster masses given by 
Horner, Mushotzky \& Scharf (1999) is
\begin{equation}
M_{200}=8.2\times10^{14}\;M_{\odot}\;
        \left(\frac{kT}{10\;{\rm keV}}\right)^{1.48}.
\end{equation}
This yields, when combined with equation (3),
\begin{equation}
f_{\rm star}=0.028\left(\frac{kT}{10\;{\rm keV}}\right)^{-0.36}.
\end{equation}
Except the slightly lower amplitudes, both equations (6) and (8) 
are consistent with
equation (1) especially when the large uncertainties in the fitting results 
[equations (1), (2) and (7)] are considered. 
In Figure 1 we illustrate the resulting 
gas fraction $f_{\rm gas}=0.16-f_{\rm star}$  using 
the stellar mass fraction of equation (6). It turns out that
the derived gas fraction of rich clusters from $f_{\rm star}$ 
roughly agrees with that determined from the X-ray observations.

\subsection{Basic equations}

The dark matter distributions in groups and clusters are assumed to 
be unaffected by the formation of galaxies and follow the universal
density profile (Navarro, Frenk \& White 1995; NFW):
\begin{equation}
\rho_{\rm DM}(r)=\frac{\delta_{\rm crit}\rho_{\rm crit}}
                 {(r/r_s)(1+r/r_s)^2}, 
\end{equation}
where $\delta_{\rm crit}$ and $r_s$ are the characteristic density and length, 
respectively, and $\rho_{\rm crit}$ is the critical density of the universe.
The virial mass is defined as 
$M_{\rm vir}=4\pi r_{\rm vir}^3\Delta\rho_{\rm crit}/3$, and  
related to the virial temperature $T_{\rm vir}$ through 
the cosmic virial theorem equation (4). 
Following Bryan (2000), we adopt the fitting formula of the 
$M_{\rm vir}$-$c$ relation suggested by numerical simulations:  
$c=8.5(M_{\rm vir}h/10^{15}\;M_{\odot})^{-0.086}$, where $c=r_{\rm vir}/r_s$ is
the concentration parameter. Since the temperature in all 
the empirical formulae including the $f_{\rm star}$-$T$ relation is 
the so-called spectral temperature $T_{\rm s}$ rather than the
virial temperature $T_{\rm vir}$, an appropriate connection  
between $T_{\rm s}$ and $T_{\rm vir}$ should be established in order
to make a meaningful comparison to observations.
Using hydrodynamic cluster simulations, Mathiesen \& Evrard (2001) have 
recently conducted an extensive comparison of different measures of the 
intracluster gas temperature including the spectral temperature, the
emission-weighted temperature and the mass-weighted (virial) temperature.
In particular,  it has been shown that $T_{\rm s}$ is generally lower 
than $T_{\rm vir}$ for clusters,
depending on the bandpass used for spectral fitting. 
We adopt the best-fit relation between  $T_{\rm s}$ and $T_{\rm vir}$ 
in the 2.0-9.5 keV band to proceed our numerical computation:
$T_{\rm vir}=0.91T_{\rm s}^{1.10}$. We have also tested another fitting
formula in the 0.5-9.5 keV band,  $T_{\rm vir}=1.17T_{\rm s}^{1.00}$,
and found that our main theoretical predictions remain unchanged. 
We will omit the subscript `s' in the spectral temperature. Whenever
our theoretical predictions are compared with X-ray observations, 
we always work with the spectral fitting temperature.

For simplicity, we assume that 
the gas is dissipationless before galaxy formation so that the gas 
traces the underlying dark matter distribution and satisfies 
the equation of hydrostatic equilibrium
\begin{eqnarray}
\frac{dP^0}{dr}=-\rho^0_{\rm gas}(r)\frac{GM_{DM}(r)}{r^2};\\
\rho^0_{\rm gas}(r)=f_b\rho_{\rm DM}(r).
\end{eqnarray}
We use the superscript `0' to denote the quantities before the formation
of galaxies in the system. The temperature profile can be obtained by
straightforwardly solving the above equations with the boundary 
restriction $T^0(r\rightarrow\infty)\rightarrow0$:
\begin{equation}
kT^0(r)=kT^* \frac{r}{r_s}\left(1+\frac{r}{r_s}\right)^2
      \int_{r/r_s}^{\infty}\frac{(1+x)\ln(1+x)-x}{x^3(1+x)^3}dx,
\end{equation}
where $kT^*=4\pi G \mu m_p \delta_c \rho_{\rm crit} r_s^2$ and 
$\mu=0.59$ is the mean molecular weight. Note that unlike Bryan (2000),
we did not choose the pressure-free external boundary condition 
$P^0=0$, i.e. $T^0=0$ at $r=r_{\rm vir}$, which would lead the entropy 
distribution $S^0=\ln T^0/(n_e^0)^{2/3}$ to turn over near the boundary.
While our boundary restriction $T^0(r\rightarrow\infty)\rightarrow0$
seems to be more natural, the validity of equation (10)
beyond $r_{\rm vir}$ becomes questionable. In this regard, our requirement of
$T^0(r\rightarrow\infty)\rightarrow0$ should not be taken too literally,
and the key point here is that $T^0(r)$ should be   
a decreasing function of radius near $r_{\rm vir}$ and tend towards 
zero when $r\gg r_{\rm vir}$.

Galaxy formation is equivalent to removing the central region of radius
$r_0$ an amount of gas to $M_{\rm star}=f_{\rm star}M_{\rm vir}$:
\begin{equation}
M^0_{\rm gas}(r_0)=0.042\left(\frac{kT}{10\;{\rm keV}}\right)^{-0.35} 
                   M_{\rm vir}.
\end{equation}
The remaining gas is redistributed by the conservation of gas mass
\begin{equation}
\int_{r_0}^{\bar{r}} 4\pi\mu_e m_p n^0_e(x) x^2 dx =
\int_{0}^{r} 4\pi\mu_e m_p n_e(x) x^2 dx,
\end{equation}
where $\mu_{\rm e}=2/(1+X)$, and $X=0.768$ is the hydrogen mass fraction 
in the primordial abundances of hydrogen and helium. Namely, 
the central region of radius $r$ is refilled with the gas which is originally
distributed at larger radii between $r_0$ and $\bar{r}$. 
This process transports the gas at $\bar{r}$ with entropy 
$S^0(\bar{r})=\ln T^0(\bar{r})/[n_e^0(\bar{r})]^{2/3}$ to a new position
$r$ by the conservation of entropy
\begin{equation}
\frac{T(r)}{[n_e(r)]^{2/3}}=
\frac{T^0(\bar{r})}{[n^0_e(\bar{r})]^{2/3}}.
\end{equation}
The original position $\bar{r}$ can be thus fixed by combining the above 
two equations.
Eventually, the saturated configuration should satisfy the simple argument
\begin{equation}
M_{\rm gas}(r_{\rm vir})+M_{\rm star}(r_{\rm vir})=f_bM_{\rm vir}.
\end{equation}
Once the equation of state (equation [15]) and the saturated 
configuration (equation [16]) are specified, we can obtain the
newly established gas distribution by solving the following
equations:
\begin{eqnarray}
\frac{dP_e}{dr}=-\frac{GM_{DM}(r)}{r^2} \mu m_p 
              \left(\frac{P_e}{kT}\right);\\
\frac{dM_{\rm gas}}{dr}=4\pi \mu_e m_p r^2\left(\frac{P_e}{kT}\right),
\end{eqnarray}
in which $P_e=n_ekT$. The resulting gas density, temperature and entropy 
distributions are shown in Figure 2 - Figure 4 for various  
temperatures ranging from 0.5 keV to 14 keV. It appears 
that both the cores in the gas density profiles and the entropy 
floors are created in the centers as a result of galaxy formation 
which has consumed the central gas by converting it into stars. 
The effect is more 
significant in lower temperature systems than in higher ones, 
as was naturally expected.  On the other hand, the conservation
of entropy during the formation of galaxies leads to a remarkable
increase of the gas temperature towards the centers of groups/clusters.
Although radiative cooling has not been included in the above 
treatment, the significantly raised gas temperature in the central
regions can help to increase the cooling time which 
scales as $T^{1/2}$. It thus remains to be an interesting issue of 
whether the combined effect can resolve or release the cooling flow 
crisis claimed by a number of recent 
observations with {\it Chandra} and {\it XMM} (e.g. Fabian et al. 2001).
For the latter, the most compelling evidence comes from  
the X-ray spectra obtained towards the central regions in clusters 
which show a remarkable lack of emission lines from gas with $T<1$ keV
(Peterson et al. 2001). One of the suggested mechanisms is that 
there is an additional transport mechanism that can 
reheat the cool gas back to the hot phase. In this regard, 
the formation of galaxies in the central regions which simultaneously 
raises the temperature of the gas may serve as a natural heating
mechanism. Of course, further work should be done to test this speculation.

\placefigure{fig2}
\placefigure{fig3}
\placefigure{fig4}

\section{Comparison to observations}

\subsection{Radial distribution of gas fraction}

Within the framework of the GG model
a significant amount of gas in the central
regions was converted into galaxies. This would inevitably lead to an 
increasing gas fraction $f_{\rm gas}$ with group/cluster radius, 
which should be directly testable by current X-ray observations. 
For this purpose, we first extract a sample of 163 clusters, which contain 
176 data points of $f_{\rm gas}$ measured at different radii, 
from an X-ray image deprojection analysis made by White, Jones \&
Forman (1997). Note that for some of the clusters 
the X-ray temperatures $T$ were estimated by the empirical 
velocity dispersion-temperature correlation.
The data points are now divided into three subsamples according to 
temperature, which contain 42 ($T>6$ keV, $\langle T\rangle=7.99$ keV), 
98 ($3$ keV $\leq T\leq6$ keV, $\langle T\rangle=4.32$ keV) 
and 36 ($T<3$ keV, $\langle T\rangle=2.36$ keV) clusters, 
respectively, and each subsample is 
properly binned for illustrative purpose. In Figure 5 
the radial variations of the observed $f_{\rm gas}$ are compared with
the predictions for three choices of $T=8$, $4.3$ and $2.4$ keV.
Next, we choose the two group samples of Mulchaey et at. (1996) and
Hwang et al. (1999) used in the construction of 
the $f_{\rm star}$-$T$ relation (see equation [1]) to demonstrate 
the radial variation of the gas fraction in low-temperature systems.
These two samples contain 21 groups with temperature ranging from
0.69 keV to 3.38 keV, and the mean temperature is 
$\langle T\rangle=1.32$ keV. The binned data are displayed in
Figure 5. It is immediately clear from Figure 5 that
there is a fairly good agreement between the model predictions and 
the observations.  The lowest temperature subsample 
($\langle T\rangle=2.36$ keV) of  White et al. (1997) is an exception, 
exhibiting a somewhat lower amplitude than the theoretical expectation.
Perhaps, this can be regarded as the most natural 
explanation of the monotonically increasing  gas fraction with radius 
seen in many X-ray clusters although a detailed analysis should be 
made for each individual case.

\placefigure{fig5}

\subsection{X-ray surface brightness profile: slope and core radius}

The X-ray surface brightness profile $S_{\rm x}(r)$ can be computed
straightforwardly in terms of 
\begin{equation}
S_{\rm x}(r)=\frac{1}{2\pi(1+z)^4} \int_r^{r_{\rm vir}}\epsilon(T,n_e)
            \frac{RdR}{\sqrt{R^2-r^2}},
\end{equation}
where $z$ is the cluster redshift, and
$\epsilon(T,n_e)$ is the emissivity which is computed by the 
Raymond \& Smith (1977) model with a metallicity of $Z=0.3Z_{\odot}$.
The resulting $S_{\rm x}(r)$ in the 0.5-2.0 keV band for nearby clusters
($z=0$) are plotted in Figure 6 for a set of
11 groups/clusters with temperatures ranging from 0.5 to 14 keV.
It appears that the overall X-ray surface brightness profiles resemble 
the conventional $\beta$ model in shape. In order to facilitate 
a comparison with the existing X-ray imaging measurements of 
groups/clusters,  we fit the predicted X-ray surface brightness profiles 
to the $\beta$ model characterized by 
$S_{\rm x}(r)/S_{\rm x}(0)=(1+r^2/r_c^2)^{(-3\beta+1/2)}$, and work out
the slope and core radius parameters, $\beta$ and $r_c$. 
Since the determinations of  $\beta$ and $r_c$ depend somewhat
on the extension of the X-ray surface brightness profile if the
the available data extend only out to two or three times the core
radius, we truncate each $S_{\rm x}(r)$ at the maximum radii set 
by the X-ray surface brightness limits
$S_{\rm limit}=2\times10^{-14}$ erg s$^{-1}$ arcmin$^{-2}$ cm$^{-2}$ 
and $S_{\rm limit}=2\times10^{-15}$ erg s$^{-1}$ arcmin$^{-2}$ cm$^{-2}$ 
in the 0.5-2.0 keV band, respectively. These two values have 
approximately covered the {\sl ROSAT} limits used in current 
observations of clusters (e.g. MME). It turns out  that all the predicted
X-ray surface brightness profiles can be nicely fitted by the 
$\beta$ model, and the best-fit $\beta$ and $r_c$ values 
for the two flux limits and different choices of redshifts 
are plotted against the X-ray temperature in Figure 7 and Figure 8, 
respectively.

\placefigure{fig6}
\placefigure{fig7}
\placefigure{fig8}

For a comparison with real observations, we extract a sample
of 93 clusters and 36 groups from the literature in which 
the best-fit $\beta$ and $r_c$ values are explicitly provided.
The major sources of references include:   for clusters,  
Rizza et al. (1998), EF, MME,  
Arnaud \& Evrard (1999), Neumannn \& Arnaud (1999),  
Vikhlinin, Forman \& Jones (1999) and Jones \& Forman (1999);
for groups, Mulchaey et al. (1996), Dahlem \& Thiering (2000),  
Helsdon \& Ponman (2000) and Takahashi et al. (2000).
In most cases, the fittings were performed by excluding
the central cooling components, which is compatible with our
model without the inclusion of the radiative cooling. 
The corresponding spectral temperature data are taken from 
the compilation of Wu et al. (1999) and Xue \& Wu (2000).

The theoretically predicted and observationally determined slope and 
core radius parameters for different X-ray temperatures
are compared in Figure 7 and Figure 8, 
respectively. Regardless of the large dispersion in the observationally
determined $\beta$ and $r_c$ especially for groups, 
the trend of the slowly increasing $\beta$ and $r_c$ with $T$ is clearly
seen, which is consistent with our predictions. In particular,
the predicted slope parameter varies  from 
$\beta\approx0.3$ for groups to $\beta\approx0.9$ for very rich clusters,
in excellent agreement with current observations. 
It should be pointed out that the trend towards shallowed surface
brightness profiles in low-temperature systems can be 
attributed to an  observational selection effect rather than 
a property intrinsic to groups/clusters. 
This can be clearly seen from the overall predicted X-ray surface 
brightness profiles for systems with different virial masses.
Such a property differs significantly from the preheating
model in which the flattening of the $S_{\rm x}$ occurs in a low-temperature
system because the preheated gas becomes harder to be trapped in the 
shallow gravitational potential well  
(see Cavaliere et al. 1997).  We have also plotted
in Figure 7 the slope parameters expected in two extreme situations
for comparison (cf. Appendix): 
(1)The isothermal model in which the gas is driven by 
purely gravitational shocks and (2) the isentropic model in which 
the gas is preheated and then collapses adiabatically into groups/clusters. 
The former may correspond to rich clusters, while the latter may be
applied to groups under the preheating hypothesis. 
These two extreme cases give approximately rise to the upper and lower
bounds to the $\beta$ values.

Whether the increasing slope parameters $\beta$ of
X-ray surface brightness profiles from groups to cluster seen in 
current observations is purely an observational selection effect or  
a property intrinsic to groups/clusters constitutes a crucial
test for the GG model. 
There are at least two major uncertainties in the current $\beta$ model
fits of the X-ray surface brightness profiles of groups: 
Firstly, for most groups a single $\beta$ model does not provide 
an adequate description of the observed data, and the fit quality 
depends sensitively on the modeling of  central galaxies, 
cooling flows or AGNs (e.g. Helsdon \& Ponman 2000); Secondly, 
the X-ray emission of most groups remains undetectable 
outside radii of typically $\sim 200$--$400$ kpc due to the limitation 
of current detector sensitivity 
(e.g. Ponman et al. 1996;  Mulchaey et al. 1996; Helsdon \& Ponman 2000), 
while the virial radii of galaxy groups 
with temperature $T\approx1$ keV are about $1$ Mpc.
It seems that current X-ray observations have only probed 
the central regions other than the whole groups. 
It is hoped that the high sensitivity measurements with the {\sl Chandra}
X-ray Observatory can provide a robust constraint on the slopes 
of X-ray surface brightness profiles for groups and poor clusters
especially at large radii.

\subsection{X-ray luminosity - temperature relation}

We now come to the global properties of groups and clusters. 
The first test that Bryan (2000) chose for the GG model is the 
X-ray luminosity-temperature relation.  While the predicted
$L_{\rm x}$-$T$ relation is essentially consistent with 
the observed data on cluster scales, there is an apparent disagreement
below $T\approx1$ keV. This could be due to the small 
sample of poor clusters and groups and/or the observational selection effect
of the finite spatial extensions of X-ray emission 
set by the X-ray flux-limited surveys.
For the latter possibility, we recall that the correction for lost
flux falling out the detection aperture has been already made 
in the computation of $L_{\rm x}$ for clusters. 
The situation for the groups with $\beta\leq1/2$
is complicated. In most cases, no aperture correction is made 
in order to avoid the divergence of total X-ray luminosity
(e.g. Ponman et al. 1996; Helsdon \& Ponman 2000).
It is thus possible that the X-ray luminosity excess on group scales 
predicted by Bryan (2000) can be reduced simply by excluding the
X-ray emission outside the detection aperture implied by
the flux limit $S_{\rm limit}$.
We now re-examine the $L_{\rm x}$-$T$ relation 
using the catalog of X-ray groups and clusters 
compiled by Wu et al. (1999) and Xue \& Wu (2000). The updated
catalog contains 57 groups and 192 clusters whose X-ray temperatures and
luminosities are both available. Furthermore, we only account for the X-ray
emission within the apertures set by the X-ray surface brightness limits 
$S_{\rm limit}=2\times10^{-14}$ erg s$^{-1}$ arcmin$^{-2}$ cm$^{-2}$ 
and $S_{\rm limit}=2\times10^{-15}$ erg s$^{-1}$ arcmin$^{-2}$ cm$^{-2}$ 
in the 0.5-2.0 keV band, respectively. It appears that 
the evaluation of the total X-ray luminosity of clusters 
is almost unaffected by these aperture limits because the 
corresponding cutoff radii are close to or beyond the virial radii.
For the latter, we simply truncate the clusters at their virial radii.   
Again, we use the Raymond \& Smith (1977) model with 
a metallicity of $Z=0.3Z_{\odot}$ to calculate the X-ray luminosity 
and plot the resulting $L_{\rm x}$-$T$ relation in Figure 9. 
It turns out that the model matches nicely the observed data
over entire temperature range.

\placefigure{fig9}

\subsection{Entropy distribution}

The last check is the distribution of the central entropy $S$ measured at
$0.1r_{200}$ against the X-ray temperature. 
We use the updated measurements
of $S(0.1r_{\rm vir})$ by Ponman et al. (1999), 
Lloyd-Davies, Ponman \& Cannon  (2000) and Xu, Jin \& Wu (2001),
which contain 67 data points ranging from $T\approx0.5$ keV to 
$T\approx14$ keV.  We demonstrate the predicted and measured 
$S(0.1r_{\rm vir})$ in Figure 10, which reinforces the finding of 
Bryan (2000) that the predicted $S$-$T$ is consistent with
the observed one. A larger sample  especially at low temperature end
will be needed for a critical test of the model prediction.

\placefigure{fig10}

\section{Discussion and conclusions}

Using various large data sources of X-ray groups and clusters 
in the literature, we have made an extensive examination of 
the properties of the galaxy formation-regulated 
gas evolution model in groups and clusters, proposed recently
by Bryan (2000) as an alternative explanation of the entropy excess 
and the steepening of the X-ray luminosity-temperature relation in 
groups and clusters. This model is based on one empirical 
correlation between the stellar mass fraction $f_{\rm star}$ and temperature
$T$ and one hypothesis that the lowest entropy gas in the central
regions of groups and clusters was converted into stars.
The reliability of the $f_{\rm star}$-$T$ relation 
has been justified by other observational facts such as
the increasing ratio of gas mass to stellar mass with the X-ray
temperature (David et al. 1990), the increasing gas fraction with 
the X-ray temperature (Figure 1) and 
the positive correlation between the mass-to-light ratio and 
the optical luminosity in clusters (Girardi et al. 2000). 
This suggests a simple scenario that galaxy formation was less
efficient in rich clusters than in poor clusters and groups. 
As for the working hypothesis, the formation of galaxies in the central 
regions of groups and clusters consumed the central gas and created
a gas cavity, which acts as a trigger for an inward gas flow or diffusion. 
As a consequence, the flat gas core and entropy floor 
developed in the centers, together with  an
increasing gas fraction with radius. 
These predictions are in quantitative agreement with the internal structures
of intragroup/intracluster gas revealed by current X-ray observations 
such as the slope and core distributions characterized by the 
conventional $\beta$ model, the radial variations of the gas mass fraction,
and the central entropy distributions measured at $0.1r_{200}$.
Another success of the model is the reproduction of the 
X-ray luminosity-temperature relation over a broad temperature range
from $T\approx0.5$ keV to  $T\approx20$ keV. 
Overall, it seems that this simple model without additional
heating can satisfactorily account for the existing 
X-ray observations.

While both the GG model proposed by Bryan (2000) without additional 
heating and the prevailing preheating model can equally explain 
the current X-ray observations, there are a number of essential 
differences which can be used to distinguish between the two
scenarios. For example, 
the GG model admits the universality of the baryon fraction 
from groups to rich clusters, while in the preheating scenario 
the baryon fraction
increases with temperature because a considerable amount of
gas may still reside outside of the virial radii as a result of
preheating especially in poor clusters and groups. A precise 
measurement of the total baryon (stars + gas) fractions in groups
and clusters can provide a critical test for the models.   
Alternatively, the X-ray imaging measurements of the surface brightness 
profiles of poor clusters and groups are probably the simplest 
way to disentangle the issue: the GG model predicts that the slope of 
the X-ray surface brightness profile at large radii remains roughly 
unchanged from groups to clusters, and the increasing slope with 
temperature seen in current X-ray observations is purely selection
effect due to the limitation of detector sensitivity and background noise. 
On the contrary, within the framework of the preheating scenario
the X-ray surface brightness profiles are intrinsically
shallower in groups and poor clusters than in rich clusters,
insensitive to observational sensitivity. The high sensitivity
X-ray observations of groups
with advanced detectors like {\sl Chandra} should be able to
set robust constraints on the slope parameters of the X-ray surface 
brightness profiles in the outer regions of groups.

Apparently, the GG model still has its own problems that need to be resolved
in the future.  First, aside from the robustness problem of the empirical 
relation between stellar mass fraction and X-ray temperature established based
on a small sample of local groups and clusters,  it is unclear as to 
how this relation evolves with redshift. 
The cosmological applications are largely limited 
if evolutionary effect is not incorporated into the present model.
It is unlikely that there is a significant evolution of groups
and clusters since $z\sim1$ if the inward gas flow or diffusion 
due to galaxy formation in the central regions ended 
at the early phase of structure formation. 
Of course, this depends on how fast the process of inward gas flow or 
diffusion may take. A related question is: Can one actually observe 
the process at high-redshift groups and clusters ?
Second, the current $f_{\rm star}$-$T$ relation indicates that nearly all 
the gaseous materials in less massive groups and giant galaxies have been 
converted into stars. If this occurred in the early stage of structure
formation, there would be no gas left in small systems. How can one 
explain the existence of a considerably large fraction of hot gas in
clusters if they formed by gravitational aggregation of individual 
low-mass objects ?  One possibility is that the formation of galaxies 
and the inward gas flow or diffusion in less massive systems proceeded rather 
slowly and a large fraction of gas had not been converted into stars 
before they merged with other massive objects. The latter process 
can raise the gas temperature and thus reduce the efficiency of 
galaxy formation, according to the $f_{\rm star}$-$T$ relation.
A sophisticated analysis incorporated with the halo merger rates 
described by the extended Press-Schechter formalism (Lacey \&
Cole 1993) may be needed to clarify the issue. 
Third, the GG model predicts that the temperature 
profile shows a dramatic increase towards the centers and 
a significant decline at large radii. 
Although the inclusion of radiative cooling is expected to 
reduce somewhat the central X-ray temperature, whether or not this can
yield a quantitative agreement with X-ray spectral analysis  
remains unclear. It is also interesting to explore the possibility
of whether or not such a rising temperature towards the centers of
clusters can alleviate the cooling flow crisis. 
Finally, it is worth mentioning that the overall radiative cooling 
may play an important role in the removal of low-entropy gas in
the centers of groups and clusters,  giving rise to an equally good
explanation of the steepening of the $L_{\rm x}$-$T$ relation
for groups and clusters (Muanwong et al. 2001).

Overall, the GG model has seemingly proved very successful at explaining
a number of internal and global properties of the intragroup/intracluster
gas. If confirmed, this would have profound implications for our understanding
of the origin and distribution of the intragroup/intracluster gas, 
the formation history of group/cluster galaxies, the missing 
baryons in the universe, and even the formation and evolution 
of structures in the universe.

\acknowledgments
We gratefully acknowledge the valuable suggestions and comments
by an anonymous referee. This work was supported by
the National Science Foundation of China, under Grant No. 19725311
and the Ministry of Science and Technology of China, under Grant
No. NKBRSF G19990754.              

\clearpage

\appendix

In this Appendix we discuss the gas distributions in two extreme cases: 
the isothermal and isentropic models. The former may work in rich
clusters where the gas was mainly heated by gravitational shocks, while
the latter may correspond to the intragroup gas which was preheated and 
then collapsed adiabatically into the gravitational potential wells of
the groups.  A detailed discussion about the X-ray luminosity-temperature 
relations in the two models can be found in Balogh et al. (1999). 
Here, we concentrate on the radial gas density profiles and use a 
new prescription suggested recently by Eke, Navarro \& Steinmetz (2001)
for the assignment of the collapse redshift of dark halos. Again, 
we adopt the NFW profile for a virialized dark halo, and determine 
its virial temperature in terms of equation (4). 
In order to fix the free parameters, $\delta_{c}$ or $r_s$ or $c$ 
in the NFW profile. The collapse redshift
$z_{\rm coll}$ for each halo identified at $z=0$  
is introduced through (Eke et al. 2001)
\begin{equation}
D(z_{\rm coll})\sigma_{\rm eff}(M_{\rm s})=\frac{1}{C_{\sigma}}, 
\end{equation}
where $C_{\sigma}\approx25$ for the $\Lambda$CDM model, 
$D$ is the normalized linear grow factor, for which
we take the approximate expression from Carroll, Press \& Turner (1992), 
$\sigma_{\rm eff}(M_{\rm s})$ is the so-called modulated rms linear density
at mass scale $M_{\rm s}$:
\begin{equation}
\sigma_{\rm eff}(M_{\rm s})\equiv\sigma(M_s)
         \left[-\frac{d\ln\sigma(M_s)}{d\ln M_s}\right],
\end{equation}
and $M_{\rm s}$ is defined as the mass contained within 
$r_{\rm max}=2.17r_s$ where the circular velocity reaches its maximum and
\begin{equation}
M_s=\frac{0.47M}{\ln(1+c)-c/(1+c)}. 
\end{equation}
We adopt the following parameterization of the power-spectrum 
of initial fluctuation 
\begin{equation}
P(k)=A\;k^n\;T_{\rm CDM}^2(k),
\end{equation}  
where $n$ is the primordial power spectrum and is assumed to
be the Harrison-Zeldovich case $n=1$,
and $T_{\rm CDM}(k)$ is the transfer function of adiabatic CDM model
for which we use the fit given by Bardeen et al. (1986):
\begin{equation}
T_{\rm CDM}(q)=  \frac{\ln(1+2.34q)}{2.34q} 
[1+3.89q+(16.1q^2)+(5.46q)^3+(6.71q)^4]^{-1/4}, 
\end{equation}  
where $q=(k/h\;{\rm Mpc}^{-1})/\Gamma$, and 
$\Gamma=\Omega_{\rm M}h\exp[-\Omega_{\rm b}(1+\sqrt{2h}/\Omega_{\rm M})]$ is
the shape parameter.
Once a power spectrum, $P(k)$, is specified, the mass variance becomes
\begin{equation}
\sigma^2(M)=\frac{1}{2\pi^2}\int_0^{\infty} k^2 P(k) W^2(kR) dk, 
\end{equation}  
in which $W(x)=3(\sin x-x\cos x)/x^3$  
is the Fourier representation of the window function.  
The amplitude $A$ in the power spectrum 
is determined using the rms mass 
fluctuation on an 8 $h^{-1}$ Mpc scale, $\sigma_8$, for which
we take $\sigma_8=0.93$ for our $\Lambda$CDM model.
Finally, the relationship between the NFW profile and its collapse reshift
$z_{\rm coll}$ is established if we define  a new 
characteristic density, $\tilde{\rho}_s=3M_{\rm vir}/4\pi r_s^3$, and 
set to equal the spherical collapse top-hat density 
at $z_{\rm coll}$: 
\begin{equation}
\tilde{\rho}_s=\Delta(z_{\rm zoll}) \rho_{\rm crit}(z_{\rm coll}).
\end{equation}

{\sl I. Isothermal model}
 
We first assume that the hot gas is isothermal with
electron number density $n_e(r)$ and temperature $T$, and 
is in hydrostatic equilibrium with the underlying gravitational
potential dominated by the NFW profile:
\begin{equation}
-\frac{GM_{\rm DM}(r)}{r^2}=\frac{1}{\mu m_p n_e(r)}
      \frac{d[kT n_e(r)]}{dr}.
\end{equation}
A straightforward computation yields an analytic form of the electron 
density profile (Makino, Sasaki \& Suto 1998)
\begin{equation}
n_e(r)=n_e(0)\frac{(1+r/r_s)^{\alpha/(r/r_s)}-1}{e^{\alpha}-1},
\end{equation}
where we have adopted the normalized, background-subtracted form
in order to ensure the convergence of the X-ray surface brightness, 
and $\alpha=4\pi G \mu m_p \delta_c \rho_{\rm crit} r_s^2/kT$.
In terms of the virial theorem, the $\alpha$ parameter reduces to 
\begin{equation}
\alpha=\frac{3c}{\ln(1+c)-c/(1+c)}.
\end{equation}
Here we have assumed that the specific kinetic energy 
of dark matter particles is equal to that of gas:
$\beta_{\rm spec}=\sigma^2/(kT_{\rm vir}/\mu m_p)=1$.
In turns out from the prescription of the dark halo formation
discussed above that $\alpha$ is a slowly 
decreasing function of virial mass or temperature, ranging from 
22.5 to 14 for $10^{9}M_{\odot}\leq M\leq 10^{17}M_{\odot}$.
This is in good agreement with 
the observationally determined values for clusters 
(e.g. EF;  Wu \& Xue 2000).

Once the electron density and temperature are specified
for different systems,
we can calculate the X-ray surface brightness profile using 
an optically thin, isothermal plasma emission model by 
Raymond \& Smith (1977). Finally, we fit the predicted
X-ray surface brightness profile to the $\beta$ model and 
work out the best-fit $\beta$ and $r_c$ parameters. 
It appears that the resulting $\beta$ parameter decreases slightly with 
temperature, as shown in Figure 7.
At the high temperature end of $T>10$ keV, the predicted 
slope parameter reaches $\beta\sim0.8-0.9$,  roughly consistent
with observed values for very rich clusters.

{\sl II. Isentropic model} 

The specific entropy of the gas can be conveniently defined as
\begin{equation}
S=\frac{kT}{n_e^{\gamma-1}},
\end{equation}
where $\gamma$ is the polytropic index. In the case of where the gas 
is accreted adiabatically, $\gamma=5/3$.
We solve the hydrostatic equilibrium equation by demanding that
the entropy of the gas is conserved during accretion.
We specify such a boundary condition that the electron number density and
temperature at large radii should approach asymptotically 
the background values $n_{e,b}$ and $T_b$, respectively.  
We caution that this last constraint may break down beyond virial radius
because of the failure of hydrostatic equilibrium. Therefore, this
boundary condition should only be taken to be a reasonable approximation.
The electron number density and temperature profiles are 
\begin{eqnarray}
n_{\rm e}(r)=n_{e,b}\left[1+\alpha_b\left(\frac{\gamma-1}{\gamma}\right)
                    \frac{\ln(1+r/r_{\rm s})}{r/r_{\rm s}}\right]
                    ^{\frac{1}{\gamma-1}};\\
T(r)=T_{b}\left[1+\alpha_b\left(\frac{\gamma-1}{\gamma}\right)
                    \frac{\ln(1+r/r_{\rm s})}{r/r_{\rm s}}\right],
\end{eqnarray}
where 
\begin{equation}
\alpha_b=\frac{4\pi G\mu m_{\rm p}\delta_{\rm c}\rho_{\rm crit}r_{\rm s}^2}
	  {kT_b}=
         \frac{4\pi G\mu m_{\rm p}\delta_{\rm c}\rho_{\rm crit}r_{\rm s}^2}
	  {Sn_{e,b}^{\gamma-1}}.
\end{equation}
Nevertheless, we use the background-subtracted electron density 
$\bar{n}_{\rm e}(r)=n_{\rm e}(r)-n_{\rm e,b}$ and 
temperature $\bar{T}(r)=T(r)-T_{b}$
to proceed our calculation of the corresponding X-ray emission, i.e.,
\begin{eqnarray}
\frac{\bar{n}_{\rm e}(r)}{n_{e,b}}=
\left\{\left[1+\alpha_b\left(\frac{\gamma-1}
                                            {\gamma}\right)
                    \frac{\ln(1+r/r_{\rm s})}{r/r_{\rm s}}\right]
                    ^{\frac{1}{\gamma-1}}-1\right\};\\
\frac{\bar{T}(r)}{T_{b}}=
\alpha_b\left(\frac{\gamma-1}{\gamma}\right)
                    \frac{\ln(1+r/r_{\rm s})}{r/r_{\rm s}}.
\end{eqnarray}
The key parameter that determines the shape of the X-ray surface brightness
profile $S_{\rm x}(r)$ is $\alpha_b$. 
The typical value of $\alpha_{\rm b}$ in terms of the above evaluations of 
$\delta_c$ and $r_s$ for a given halo $M_{\rm vir}$ or $T_{\rm vir}$ is 
$\alpha_b\sim 4\pi G \mu m_p \delta_c 
\rho_{\rm crit} r_s^2/kT_{\rm vir}\sim10$ for 
$10^{9}M_{\odot}\leq M\leq 10^{17}M_{\odot}$.
We have numerically obtained the profiles of $S_x(r)$ for
a number of $\alpha_b$ and fitted them to the $\beta$ model over a 
radius range of $0\leq r \leq 100r_s$. 
The resulting $\beta$ starts from $\sim0.3$ for $\alpha_b=10$
and approaches $\sim0.4$ for the extremely large values of 
$\alpha_b\sim10^4$ (see Figure 7). Such a broad range of $\alpha_b$
should cover  a broad mass range from groups to rich clusters.  
It appears that the saturated configuration of the isentropically
accreted gas in an NFW-like gravitational potential well looks
quite shallow and can be well represented by the conventional $\beta$ model
with $\beta\approx0.3-0.4$.

\clearpage

\figcaption{Gas mass fraction within $r_{500}$ versus X-ray temperature
for nearby rich clusters. A total of 55 clusters from 
Mohr et al. (1999; MME) and Ettori \& Fabian (1999; EF) are shown.
Also plotted are the expectations from the correlations between the  stellar 
mass fraction and X-ray temperature given by Bryan (2000) (solid line)
and between the virial mass and optical luminosity found by 
Girardi et al. (2000) (dotted line), in combination
with the universality of the total baryon fraction in clusters
predicted by the Big Bang Nucleosynthesis (BBN) (dashed line).
\label{fig1}}

\figcaption{Normalized gas density profiles for  
different X-ray temperatures ranging from 
0.5 keV to 14 keV. The initial profiles and 
the galaxy formation-regulated gas distributions 
are plotted by dotted and solid lines, respectively. 
\label{fig2}}

\figcaption{The same as Figure 2 but for the true temperature profiles.
\label{fig3}}

\figcaption{The same as Figure 2 but for the entropy profiles.
\label{fig4}}

\figcaption{Radial variations of the gas mass fraction.
The predicted profiles (dotted lines, from top to bottom) correspond 
to four different choices of temperatures $T=8.0$, $4.3$,
$2.4$ and $1.3$ keV.
Also shown are the observationally derived 176 data points of clusters 
(open symbols) by White et al. (1997) and 21 data points
of groups (filled stars) by Mulchaey et al. (1996) and Hwang et al. (1999) 
within different radii, which are properly binned 
according to temperature and radius. 
\label{fig5}}

\figcaption{The predicted X-ray surface brightness profiles 
in the 0.5-2.0 keV band for a set 
of 11 groups/clusters with temperatures ranging from 0.5 to 14 keV 
(from bottom to top). Dotted lines correspond to the X-ray surface
brightness limits 
$S_{\rm limit}=2\times10^{-14}$ erg s$^{-1}$ arcmin$^{-2}$ cm$^{-2}$ 
(upper) and 
$S_{\rm limit}=2\times10^{-15}$ erg s$^{-1}$ arcmin$^{-2}$ cm$^{-2}$ 
(lower), respectively.
\label{fig6}}

\figcaption{Dependence of the slope parameter ($\beta$) upon 
X-ray temperature ($T$).  A total of 127 groups and clusters
extracted from the literature are shown.
Thick and thin solid lines are the best-fit $\beta$ values 
from the GG model with and without the correction for  
maximum detectable radius, among which the three thick solid lines 
from top to bottom correspond to ($S_{\rm limit}$, $z$)=
($2\times10^{-14}$ erg s$^{-1}$ arcmin$^{-2}$ cm$^{-2}$, $0$),
($2\times10^{-15}$ erg s$^{-1}$ arcmin$^{-2}$ cm$^{-2}$, $0$),
and 
($2\times10^{-15}$ erg s$^{-1}$ arcmin$^{-2}$ cm$^{-2}$, $0.2$),
respectively. Dashed line is the best-fit value from the isothermal model.
Dotted lines represent the results in  
the isentropic model for  $\alpha_b$ ranging from
$10$ to $10^4$, which covers a broad mass range from groups to 
rich clusters (see Appendix).
\label{fig7}}

\figcaption{Comparison of the theoretically predicted core radii
and the best-fit values from X-ray observations 
for different temperatures. Virial theorem is used to assign
$r_{200}$ to each group/cluster. Thick and thin lines correspond to
the $\beta$ model fittings with and without the correction for  
maximum detectable radius. The notations are the same as in Figure 7.
\label{fig8}}

\figcaption{Comparison of the predicted bolometric X-ray 
luminosity-temperature relation and the observed data.  
A total of 57 groups and 192 clusters with available $T$ and 
$L_{\rm x}$ in the literature are shown. 
The upper and lower lines correspond to the two flux limits shown in
Figure 6, 
$S_{\rm limit}=2\times10^{-14}$ erg s$^{-1}$ arcmin$^{-2}$ cm$^{-2}$ 
and 
$S_{\rm limit}=2\times10^{-15}$ erg s$^{-1}$ arcmin$^{-2}$ cm$^{-2}$, 
respectively.
\label{fig9}}

\figcaption{Dependence of the central gas entropy  measured 
at $0.1r_{200}$ upon the X-ray temperature. The model prediction
is plotted by solid line. The observational results from 
Ponman et al. (1999; PCN),  Lloyd-Davies et al. (2000; LPC) and 
Xu et al. (2001; XJW) are clearly marked. 
\label{fig10}}

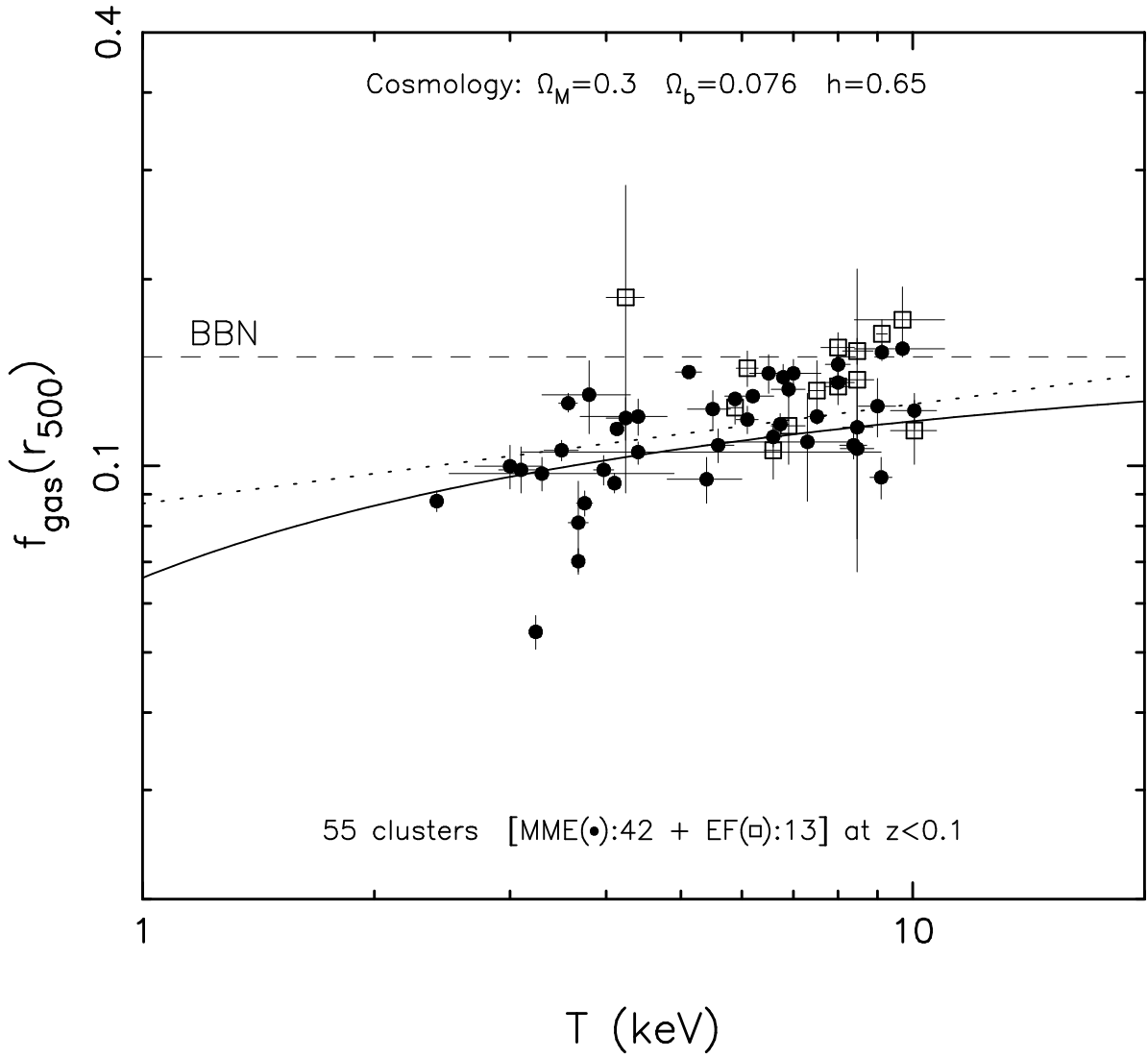
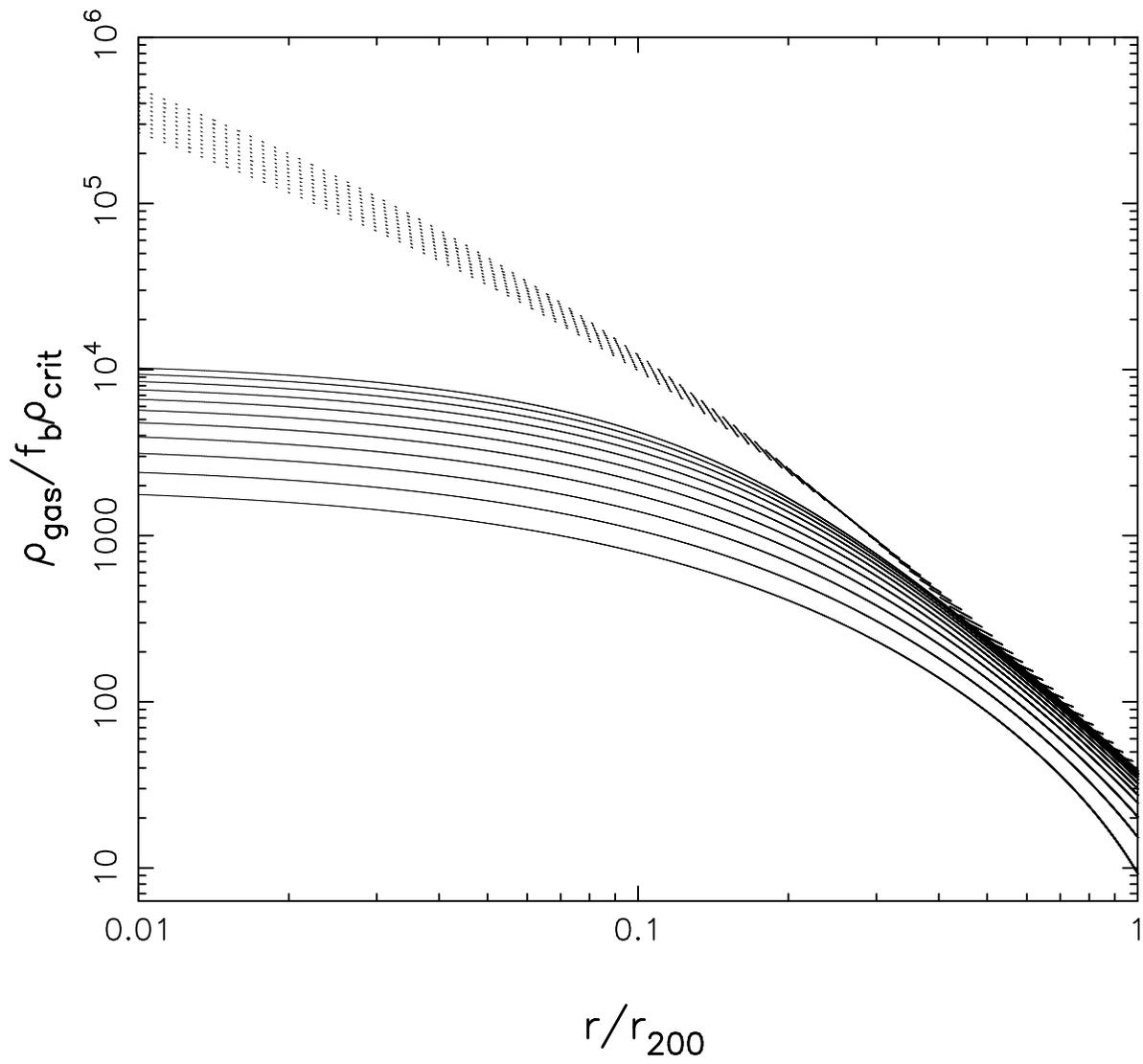
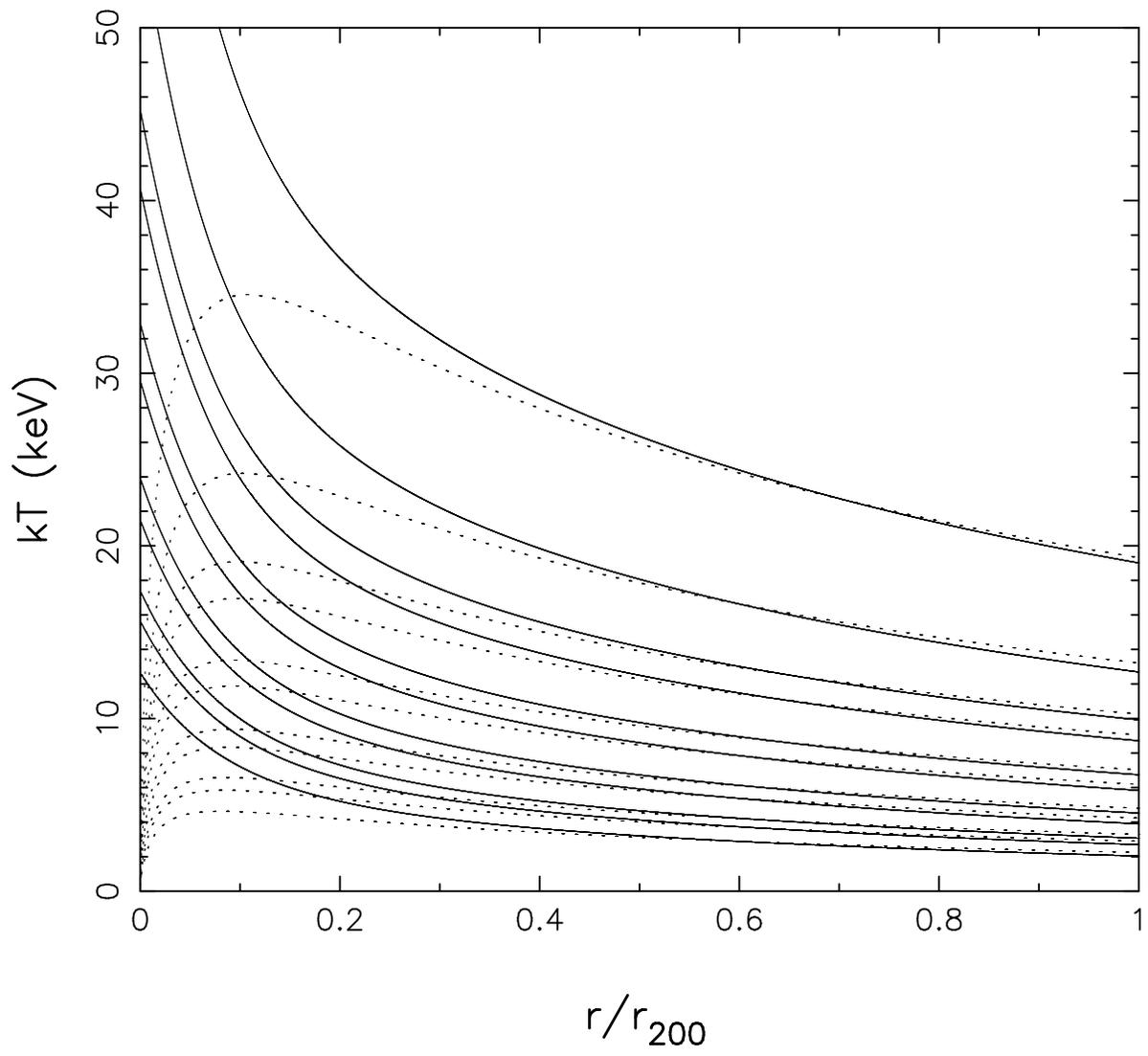
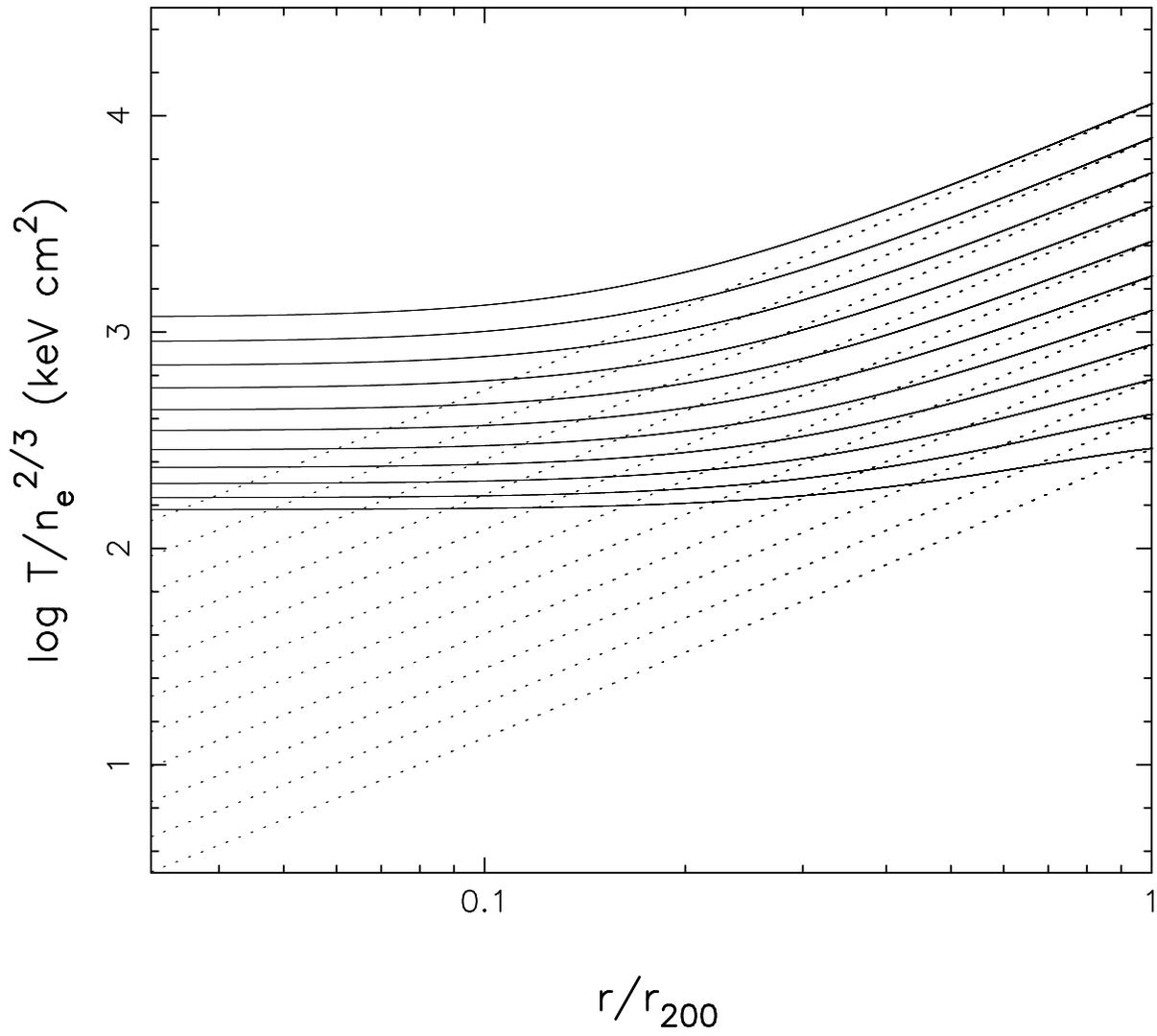
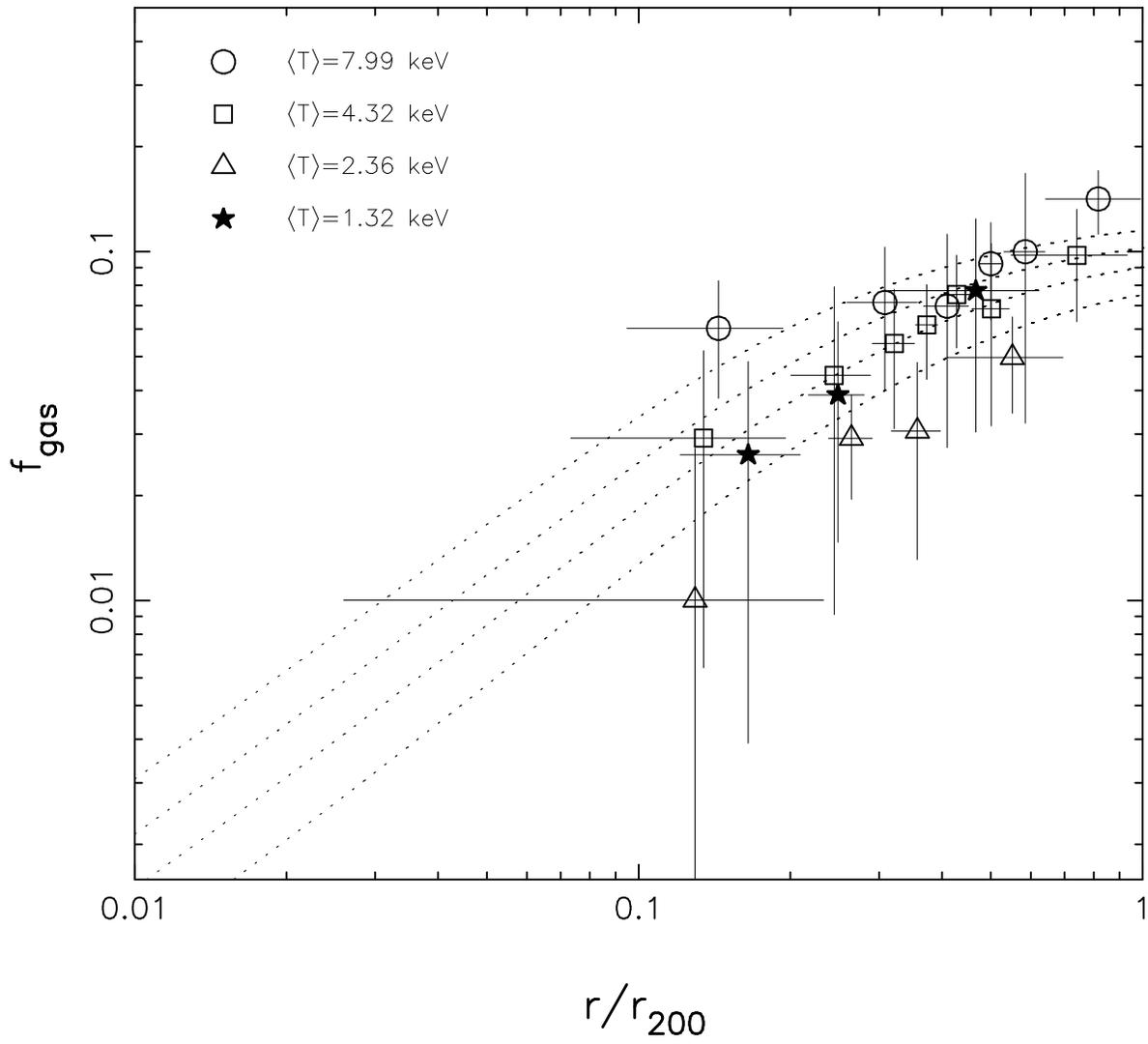
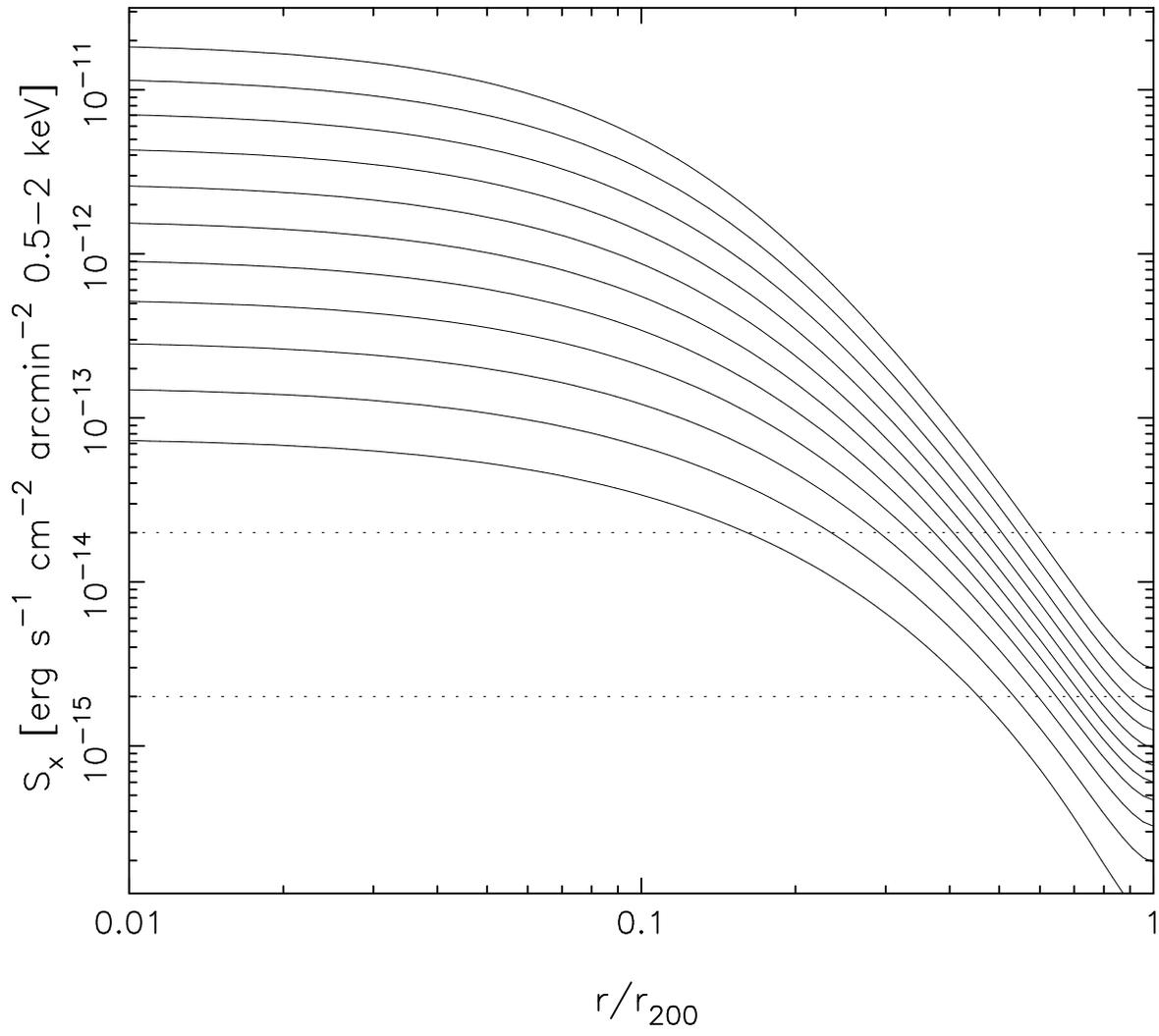
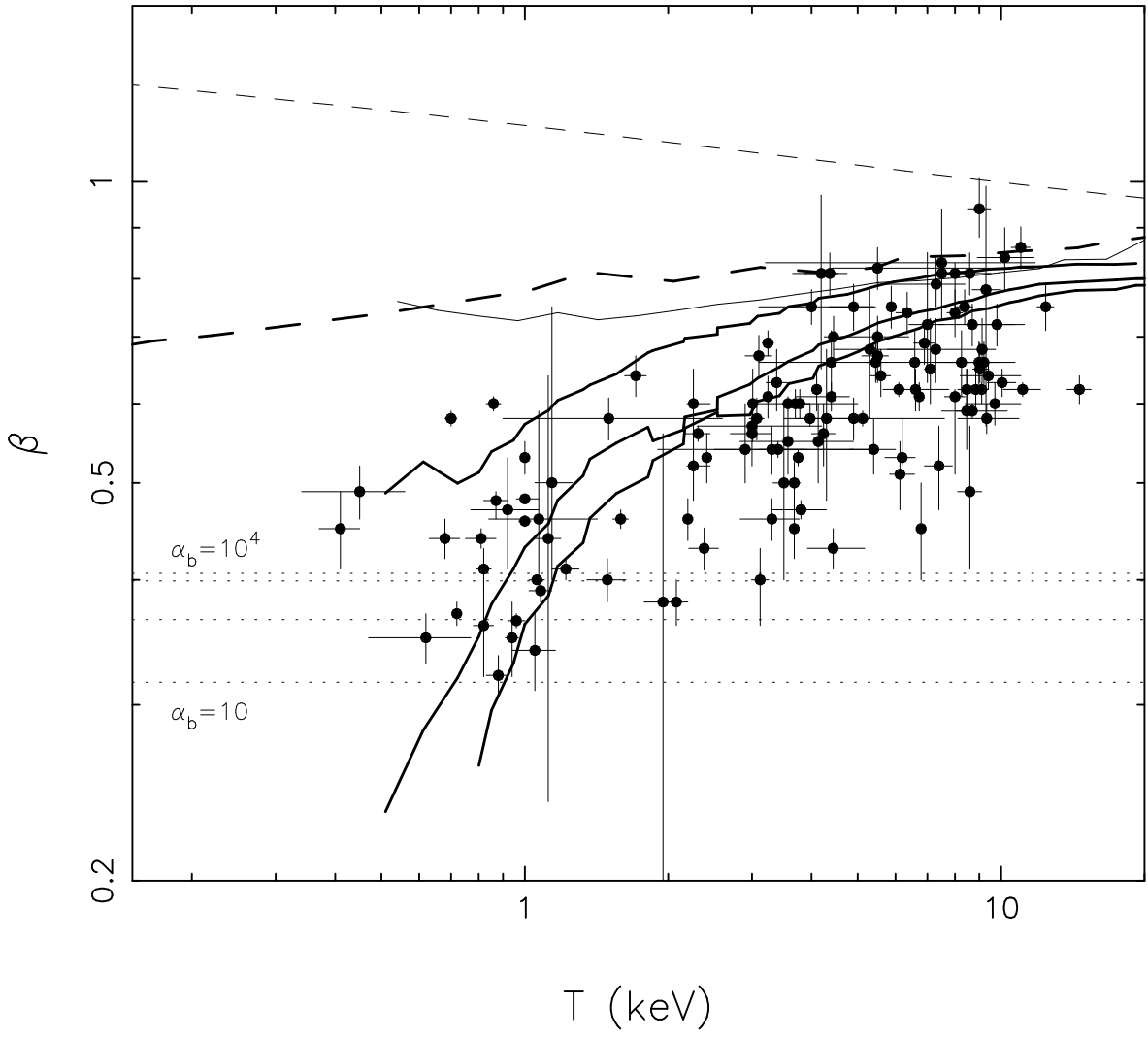
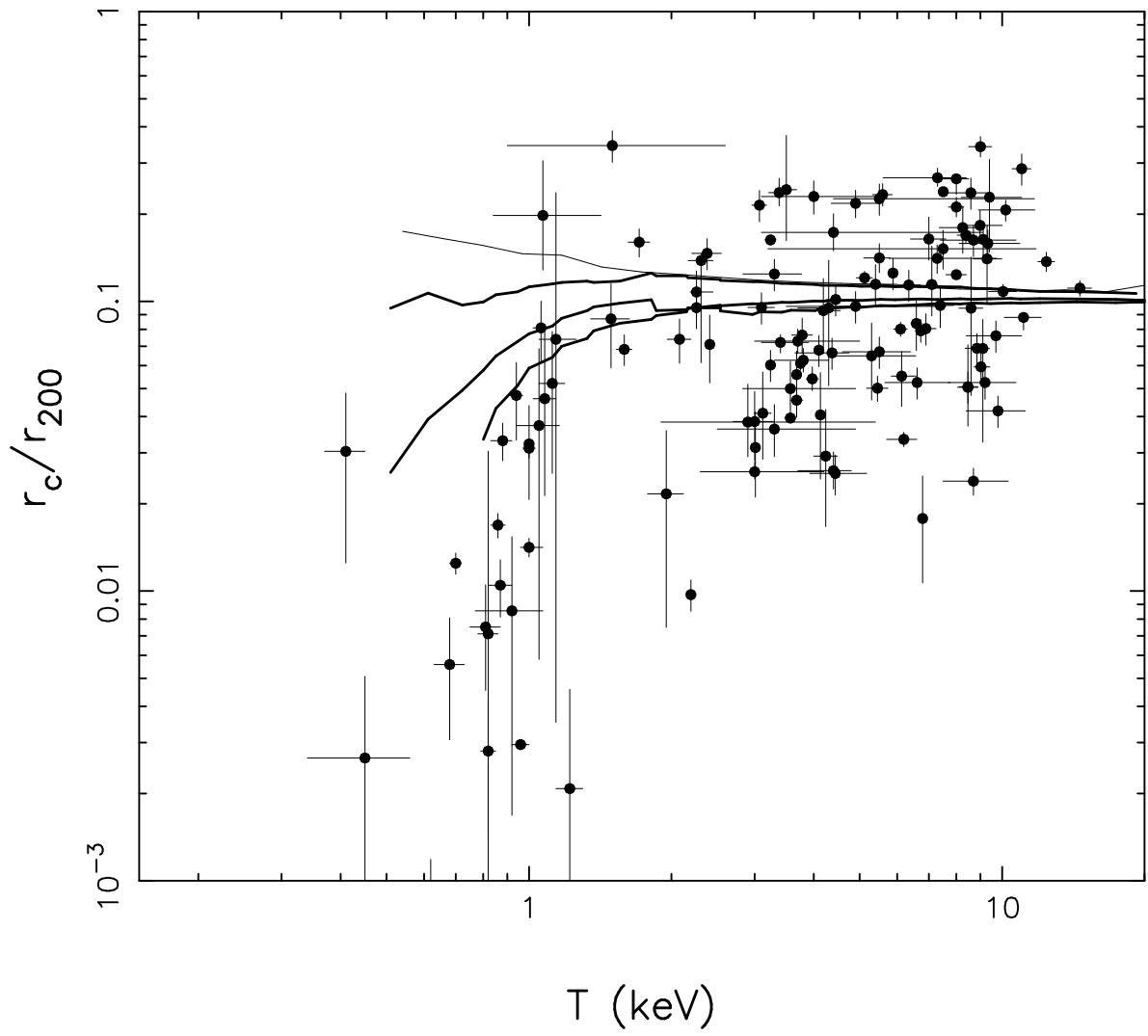
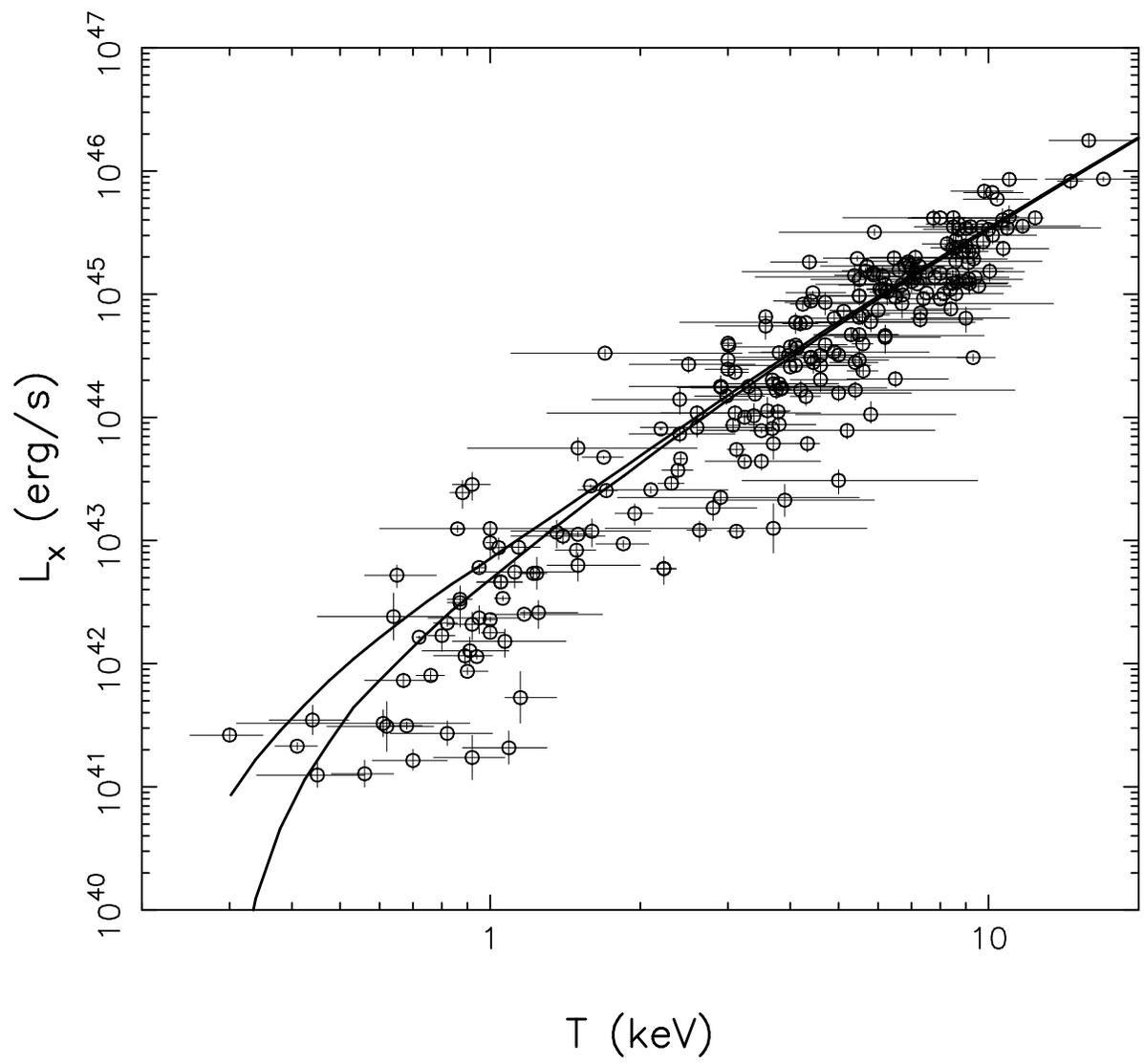
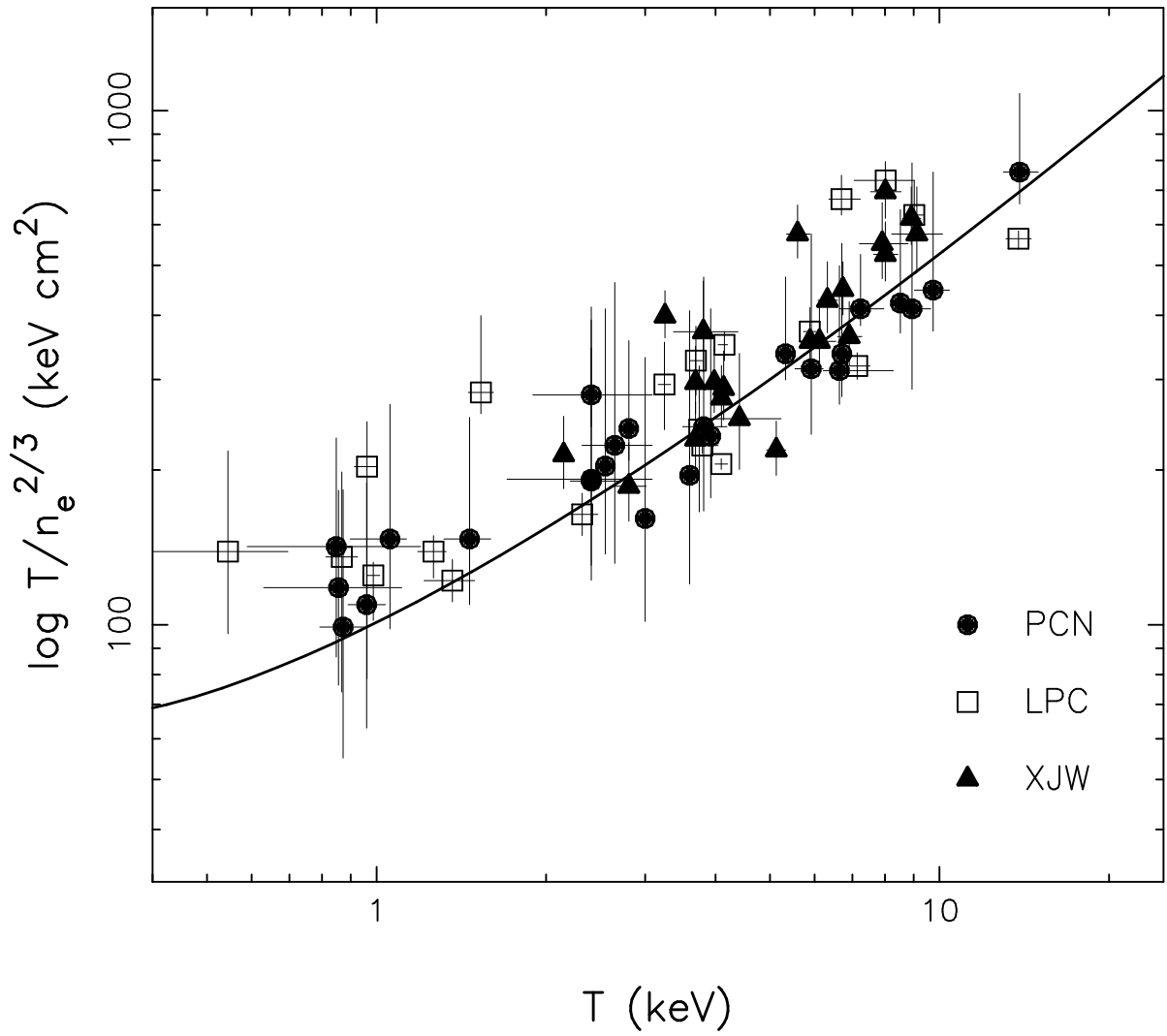
\end{document}